\documentclass[12pt]{article}
\usepackage{amssymb}
\usepackage{epsf}
\usepackage{lscape}
\usepackage[T2A]{fontenc}
\usepackage[english]{babel}
\begin{document}
\begin{center}
Comments to the paper ``\,Effect  of   Coulomb forces on the
Position of the Pole  in the Scattering Amplitude  and on Its
Resudue'' published in Phys.At.Nuclei, {\bf 73} (2010) 757
\end{center}
\begin{center}
  R. Yarmukhamedov\\
\vspace{0.5cm}
Institute of Nuclear Physics, Uzbekistan Academy of Sciences, 100214 Tashkent, Uzbekistan\\
\end{center}
\vspace{1cm}
\begin{center}
Abstract:

\hspace{0.6cm}Certain comments on the paper of Yu.V. Orlov, B. F.
Irgaziev and L. I. Nikitina published in Phys.At.Nuclei, {\bf 73}
(2010) 757 are made.
\end{center}
\hspace{0.6cm}  In  \cite{Orlov2010}, the explicit form of the
nuclear vertex constant (NVC) for the virtual decay $a\to b+c\,$
with two charged particles ($b$ and $c$) and an arbitrary orbital
momentum $l_a$  is derived for the standard effective-range
expansion $K_{l_a}(k^2)$. There, as an example, the bound and
resonance states of the lightest nuclei, including the ${\rm
{^5He}}$ and ${\rm {^5Li}}$ nuclei in the $p$-wave, are studied. It
should be noted that the similar expressions have earlier  been
derived in [2--4] independently, but in the other forms. For
example, two the similar expressions are derived in \cite{Mark2009}
for the asymptotic normalization coefficient (ANC), which is
proportional to the nuclear vertex constant up to the known
multiplicative factor \cite{Blokh1977}. One of them is valid for the
neutral case and another one is valid only for the charged case, and
the latter has no  limit when a charge of a particle (either $b$ or
$c$) tends to zero.

As it is seen from here,   at   present obtaining a correct relation
between the NVC (or ANC) and the parameters of the effective range
expansion is of great interest since it makes it possible  to
determine the aforesaid fundamental characteristic bound state of
the nucleus  $a$ in the ($b$+$c$)-configuration and the parameters
of the effective range expansion for the $bc$-scattering by
selfconsistent way.

In  work \cite{Orlov2010}, the results of work  \cite{Igam1997} is
also criticized.  In particular, in Introduction of \cite{Orlov2010}
the authors  assert  that   ``\,...a serious error of fundamental
importance was made in \cite{Igam1997}: Eq. (25) \footnote{{\it In}
\cite{Igam1997}, {\it there is the misprint in the expression (25)
(see the works \cite{Igam2008,Igam2010})}{\it ; further in }
\cite{Igam1997}, {\it    the correct expression was in reality used
(here and below the italics are made by us ).}} in \cite{Igam1997}
(the numbering of  the  formulas   in this  and in the next section
corresponds strictly to the numbering in \cite{Igam1997}), which
relates  the binding energy to the ``scattering length'' and to the
``effective range'', was written without allowance for Coulomb
interaction (!). Equation (25), which is inappropriate in the case
of charged particles, was used there to derive a formula for the
elastic-scattering amplitude \{see Eq.(23) in \cite{Igam1997}\}, ...
The ``\,hybrid''  scattering amplitude obtained in this way ({\it
determined by the expressions (23)--(25)}) does not have a pole at
the binding energy $\varepsilon_{bc}$  not allowing for Coulomb
interactions or at the correct binding-energy value
$\varepsilon_{bc}^{NC}$, which includes the Coulomb interaction....
 Thus, expressions (30) and (31) in \cite{Igam1997} for the vertex
  constant $G_l^{NC}$ ({\it $G_{bc;\,l_as_a}$ in the denotation of
} \cite{Igam1997}) characterizing the virtual decay of nucleus $a$
to two charged fragments,  $a\to d+c $, are erroneous.'' (Here and
below the numbering of    the formulas  corresponds to
\cite{Igam1997}, if another is not
 noted specially, and  the phrases inside  the quotation  marks belong to
authors of   \cite{Orlov2010}).

We  may agree with these asserts partially, namely,  that is related
to the expression (25). Nevertheless, in \cite{Igam1997} a use of
the approximated equation
  (25) (the connection equation for the bound ($b$+$c$)
state)  in the partial scattering amplitude (23) does not influence
the derivation of the expressions (30) and (31) since this equation
modifies only the first term of the denominator of the amplitude
(23) (or (24)), which does not depend in reality  on the variable of
the relative momentum $k$ (or the energy $E$). Consequently, the
result of differentiation of this denominator over the variable of
$E$ (or $k$)    does not depend on a choice of the form of the
equation (25). Unfortunately, the authors of \cite{Orlov2010} do not
pay attention to this obvious fact. Therefore, in the chosen
normalization for the Coulomb-nuclear part of the partial amplitude
of   $bc$-scattering given by the expression (19) (or (23)), the
expressions (30) and (31) connecting NVC $G_{bc;\,l_as_a}$ with the
effective radius parameter are correct. One notes that these
expressions were derived by us in two   independent ways.
Unfortunately, one cannot compare the expressions (30) and (31) of
\cite{Igam1997} with the analogous one (27) of \cite{Orlov2010}
since a result of differentiation of the denominator over the
variable $k$ (or $E$) is not presented in \cite{Orlov2010}. In
addition, the
 formula (27) derived in  \cite{Orlov2010} can  be obtained
directly from   the    analogous one of \cite{Iw1984} but derived
earlier for the ANC (see p.350 there), if one makes use of  the
known relation between the ANC and NVC \cite{Blokh1977}. Therefore,
in reality,  the formula (27)  of \cite{Orlov2010} was firstly
obtained in \cite{Iw1984}, but not in \cite{Orlov2010}, as it is
asserted by authors of \cite{Orlov2010}.

Besides, it should be noted that   the normalization for the
Coulomb-nuclear part of the partial amplitude (19) (or (23)) chosen
in \cite{Igam1997}  differs from that in
\cite{Orlov2010,Iw1984,Mark2009} by a factor of the Coulomb phase
multiplicative $e^{2i\sigma_l(k)}$. Allowance  of  this factor in
the corresponding expressions of \cite{Igam1997} results  in the
renormalization of the right hand side of the expressions (31) and
(34). In this case, the factor $K(\eta_B)$  entering in the
nominator of the right hand side of  the aforesaid expressions must
be replaced by the factor
$\Gamma^2(l_B+1+\eta_B)/(l_B!)^2D_{l_B}(-i\eta_B)$.

It should be noted that in \cite{Igam2010} the equation (25) of
\cite{Igam1997} has already been generated for charged particles
($b$ and $c$), which transforms to the   equation (25) when a charge
of the particle $b$ (or $c$) tends to zero.  Combining of the
expressions (30) and (31) with the generated equation (25) presented
in \cite{Igam2010} makes it possible to express the NVC (or ANC)
through the parameter of the `` scattering length'' and, distinction
on the similar relations (14), (17) and (18) of \cite{Mark2009},
this relation is valid both for the charged case and for the neutral
one. Therefore, this combined expression can be applied for getting
an information about the scattering data if  a value of the NVC (or
ANC) is known.

However,  for some of the specific scattering considered in
\cite{KB2007}, including the $\alpha t$-scattering too, the
additional phase analysis performed by us, where    the information
about the ``experimental'' value for the corresponding ANC  and  the
aforesaid generated equation  \cite{Igam2010} are taken into
account, shows that in \cite{Igam1997} a use of the approximation
for the effective range expansion restricting  by terms up to $k^2$
does not allow one to reproduce the corresponding shift-phase
scattering at low energies by the selfconsistent   manner. In
\cite{Igam2010}, the new results   for the modified values of the
parameters of the effective-range expansion (``scattering length''
and the ``effective range'') and   the $p$-wave phase shifts
obtained for the $\alpha t$-scattering have already  been given.

In this connection, at   present the   aforesaid expression for the
NVC (or ANC) and the corresponding   connection equation for the
bound ($b$+$c$) state have been generated by us for the effective
expansion function $K_{l_a}(k^2)$ restricting  by terms up to $k^6$.
These expressions were also applied for an analysis of the
experimental phase shift scattering considered in \cite{KB2007}.  In
particular, combining of these expressions with the known
``experimental'' values of the ANCs for  the ground and first
excited states of ${\rm {^7Li}}$  in the ($\alpha$+$t$)-channel
\cite{Igam1997} makes  it possible one to   reduce the number of the
free effective expansion parameters on two. As a result, the values
of these parameters found by this way reproduce rather well the
experimental $p$- wave phase shifts  for   $\alpha t$-scattering at
energies up to about 5 MeV. This result and the similar ones for the
other scattering considered in \cite{KB2007} will be presented for
publication in a form of  a separate  paper.

It should be noted that these expressions can also be used   for
resonant states of the nucleus $a$. For this,   the binding energy
$\varepsilon_{bc}$ (or $\varepsilon_{bc}^{NC}$) should be replaced
by -$E^{(r)}$+$i\Gamma/2$, where $E^{(r)}$($\Gamma$) is the energy
(width) of the resonant state of $a$.

The author thanks   L. D. Blokhintsev for useful discussions,   D.
Baye and the authors of \cite{Orlov2010} for the comment made about
the equation (25) of \cite{Igam1997}.


\vspace{1cm}

\begin{thebibliography} {*}
\bibitem{Orlov2010}
Yu. V. Orlov, B. F. Irgaziev, and  L. I. Nikitina,
Phys.At.Nucl.,{\bf 73} (2010)757.
\bibitem{Iw1984}
Z. R. Iwinski,   L. Rosenberg, L. Spruch, Phys.Rev. C {\bf 29}, 349
(1984).
\bibitem{Igam1997}
S. B. Igamov, R. Yarmukhamedov, Nucl.Phys.A {\bf 781}, 247 (2007).
\bibitem{Mark2009}
Jean-Mark Sparenberg, P. Capel, and D. Baye, Phys.Rev.C {\bf 81},
011601(R) (2010).
\bibitem{Blokh1977}
L.D. Blokhintsev, I.Borbely, E.I. Dolinskii, Fiz.Elem.Chastits At.
Yadra. {\bf 8}, 1189 (1977)[Sov. J. Part. Nucl. {\bf 8}, 485 (1977).
\bibitem{Igam2008}
S. B. Igamov,   and R. Yarmukhamedov, Phys.Atomic Nucl. {\bf 71},
1740 (2008).
\bibitem{Igam2010}
S. B. Igamov, R. Yarmukhamedov, Nucl.Phys.A {\bf 832}, 346 (2010).
\bibitem{KB2007}
R. Kamouni, D. Baye, Nucl.Phys.A {\bf 791}, 68 (2007)
\end{thebibliography}
\end{document}